# RF front-end design and simulation for Sub-picosecond bunch length measurement

Duan Li-Wu(段立武)[1)]   Yuan Ren-Xian(袁任贤)[1)]   Leng Yong-Bin(冷用斌)[1)]

[1]Shanghai Institute of Applied Physics, Chinese Academy of Sciences, Shanghai 201808, China

**Abstract**: Cavity Beam Length Monitor is beam length measurement detector metering ultra short bunch. We designed a RF front-end and make simulations to testify this has high signal-to-noise ratio ensuring beam length measurement precision.

## 1 Introduction

Shanghai Pilot X-ray Free Electron Laser is being establishing, to produce more short light pulse with high brightness and coherence. Bunch length of xFEL is less than one picosecond, and accuracy of bunch length measurement needs to reach sub-picosecond.

Different methods are used to measure bunch length in global world, and its resolution can reach one picoseconds but also have some disadvantage. Streak-Camera Measurement can meter beam length at 1 picosecond and its resolution is 200 femtosecond [1], but its expense is too high. Michelson interferometer can meter beam length at 100 femtosecond with resolution at 30 femtosecond, but what it meters is long-time average value [2]. Zero phasing method can meter beam length at 100 fetmosecond with resolution at 10 fetmosecond, but it needs extra deflection magnet [3].

Cavity Beam Length Monitor(CBLM) which possesses high signal accuracy could be used to meter less than sub-picosecond bunch length, and data acquisition and processing has to ensure that the systemic signal noise ratio is up to 110dB. And we design a analog front-end and testify it can achieve demand.

## 2 Bunch Measurement Principle

Bunch longitudinal distribution in FEL doesn't match Gaussian formula. But as to random bunches, their frequency distribution can be written as [4]:

$$F(w) = \int_{-\infty}^{\infty} f(t)e^{-jwt}dt = \int_{t_{min}}^{t_{max}} f(t)(1 - \frac{w^2 t^2}{2} + \frac{w^4 t^4}{24} + L)dt \quad (1)$$
$$= q(1 - \frac{1}{2}w^2 t_{rms}^2 + \frac{\beta w^4 t_{max}^4}{24})$$

In Eq(1), $t_{rms}$ represents bunch length in RMS, q is bunch electric charge, **ω** is bunch frequency, $t_{max}$ represents bunch maximum length, **β** is a form factor of bunch frequency distribution. The third section of the formula can be neglected via calculation.

While we design two different Cavity Beam Length Monitor to emit two coupling signals of different frequencies, bunch RMS length can be calculated approximately as:

$$\sigma_L^2 = \frac{c_0^2}{2\pi^2} \frac{V_2 R_1 - V_1 R_2}{V_2 R_1 f_1^2 - V_1 R_2 f_2^2}, R = f_{010}\sqrt{\frac{Z}{Q_L}\left(\frac{R}{Q}\right)_{010}} \quad (2)$$

In Eq(2), $R_i$ describes signal pickup capability of cavity, which can be calibrated in laboratory. c is velocity of light, $f_i$ is cavity coupling signals frequency.

While bunch length is shorter than c/f, the resolution of bunch length is:

Supported by National Natural Science Foudation of China(Y415321061)
Subbmitted to 'Chinese Physics C'

$$\Delta\sigma_L = \frac{c_0^2}{2\pi^2\left(f_{010,2}^2 - f_{010,1}^2\right)} \cdot \frac{1}{\sigma_L \cdot SNR_V} \qquad (3)$$

Among Eq(3), $SNR_V$ is systemic signal-to-noise ratio, and $TM_{010}$ of cavity has better signal-to-noise ratio than 112dB. While the $SNR_V$ stay fixed, the beam length and its measurement resolution are negative relative. As shown in Figure 1, the resolution can be 0.35 micrometer while bunch length is 300 micrometer; the resolution can be 3.5 micrometer while bunch length is 30 micrometer.

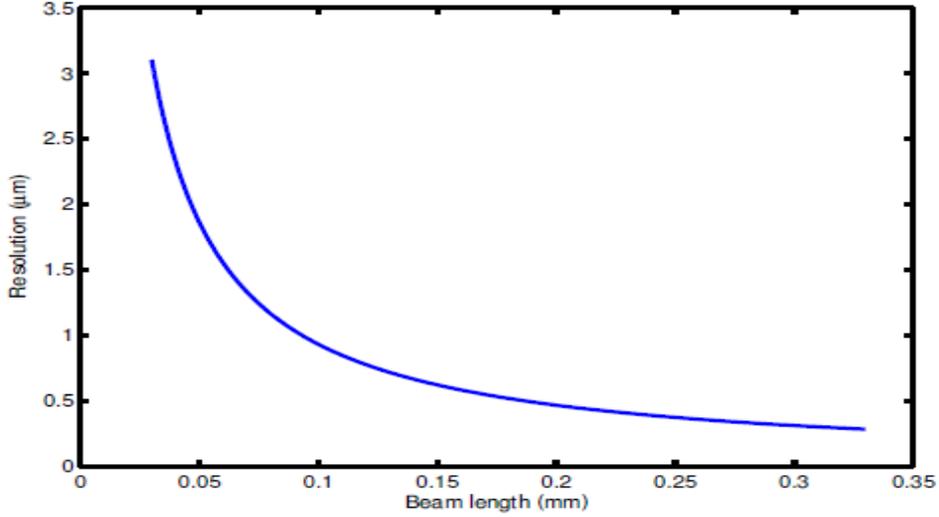

Fig 1. Relation between beam length and its resolution

**3 Analog RF Front-end Design**

Signals emitted from CBLM are about at 3 G rating and 8 G rating. Before digitalized into ADC, the signals should be down-converted to medium frequency, which is called superheterodyne. We design two down-conversion, the first is that signals of two rating are changed to 2.5 G rating, and then down-converted to tens million rating. The same crystal oscillator is used to be the local port while doing down-conversion, in order to avoid variant influence from different oscillator.

We want to ensure the systemic signal-to-noise ratio as 112dB, which makes us need a good ADC to digitalizing signals whose valid bit is upon 20 bits. It's an impossible task, we cannot find a ADC with such high valid bits in any ADC manufacturer. In order to solve this difficulty, we think out a ingenious idea. In front of signals digitalizing, we make subtraction and summation to that two channel signals. Before this, two channel signals have been adjusted to almost equal and their differential is upon 60dB smaller than quondam amplitude. Then signal after subtraction and summation are inputted to ADC. Above this, the beam length and its resolution are:

$$\sigma_L = \frac{c_0}{\sqrt{2}\pi} \cdot \sqrt{\frac{K_R - \left(1 + 2\Delta V/\Sigma V\right)}{K_R f_1^2 - \left(1 + 2\Delta V/\Sigma V\right)f_2^2}} \qquad (4)$$

$$\Delta\sigma_L = \frac{c_0^2}{2\pi^2\left(f_2^2 - f_1^2\right)} \cdot \frac{1}{\sigma_L} \cdot \frac{2\Delta V}{\Sigma V} \cdot \left(\frac{noise_{\Delta V}}{\Delta V} + \frac{2noise_{\Sigma V}}{\Sigma V}\right) \qquad (5)$$

In Eq(4), $\Delta V$ is the subtraction of two signals and $\Sigma V$ is the summation. $K_R$ is determined by the ratio of $R_1$ and $R_2$.

In Eq(5), $noise_{\Delta V}$ and $noise_{\Sigma V}$ are the noise magnitude of subtraction and summation respectively. In contrast to Eq(3), if we suppose the systemic noise stay unchangeable, we could find Eq(5) is equal to Eq(3), which means the systemic measurement resolution do not worsen.

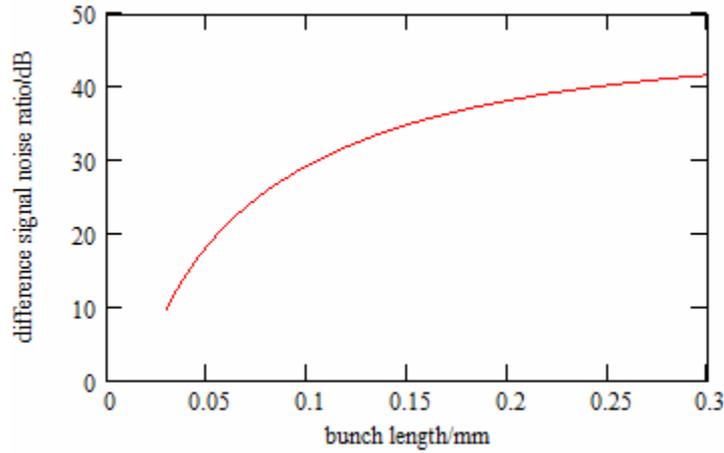

Fig 2, SNR of subtraction signal with bunch length

Fig 2 is our testing result. If beam length is 0.03 millimeter, the SNR of subtraction signal is just about 10 dB, in which systemic measurement resolution could reach 5 micrometer. Thus can widely decrease ADC's requirement.

**4 The structure of subtraction and summation**

The two channel signals are down-converted to tens million rating, and we adjust their differential is being 60dB smaller than quondam signal. Suppose one signal is $V_1$, the one is called $V_2$; then, $V_2-V_1=1‰V_1$.

How to make this subtraction and summation come true, different structure have different effects, for which the differential of $V_1$ and $V_2$ is so tiny. We design two different structure and make simulations to analyze.

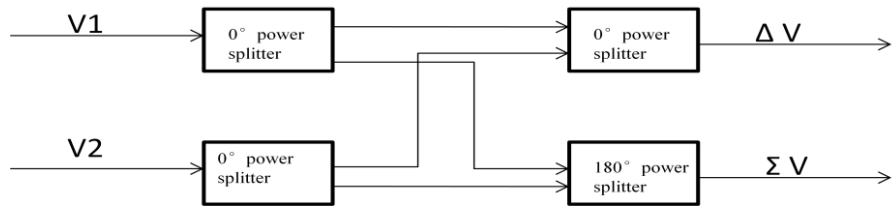

Fig 3, structure 1 of subtraction and summation

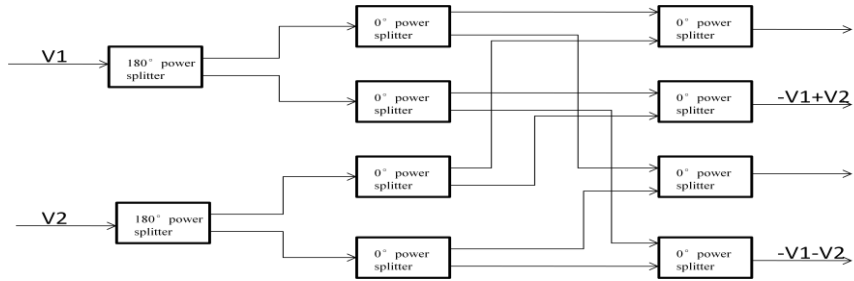

Fig 4, structure 2 of subtraction and summation

Fig 3 and Fig 4 is two different structures of subtraction and summation,. The first one is simple and the second is more complicated, but their performance is much divergent. In the first one, ΔV is equal to $V_2-V_1$ and $\Sigma V$ is equal to $V_2+V_1$ theoretically. But in fact, $\Delta V$ is much large than $V_2-V_1$, there exist coupling from the down rout. That is absolute wrong if we take $\Delta V$ as $V_2-V_1$ to do hinder compute. So we design the second structure in order to reduce the coupling from byway. In Fig 4, there are four routs: $V_1+V_2\backslash -V_1+V_2\backslash V_1-V_2\backslash -V_1-V_2$. There still exist coupling from other byways. But we can do a subtraction from $-V_1+V_2$ to $V_1-V_2$, we called it $\Delta V_2$, and we find it has less coupling influence in contrast to $\Delta V$ via experiment .

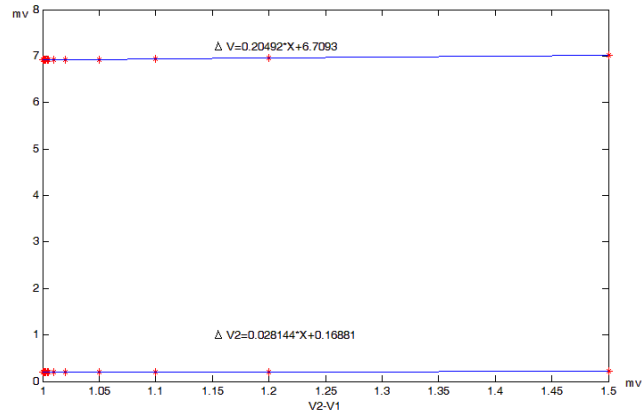

Fig 5, different $\Delta V$ of two structures

Fig 5 is the contrast about these two structures. In experement, we change the magnitude of $V_2-V_1$ from 1.001mV to 1.5 mV, and keep an account of $\Delta V$ and $\Delta V_2$. $\Delta V$ is above 40 bigger than $\Delta V_2$, in which the coupling from other byways we could imagine is so grave.

**5 Advanced Design System simulation**

Advanced Design System is a Electronic Design Automatic software developed by Agilent corporation, which has the function of circuit simulation and module exploitation. In our experiment, we use ADS software to do simulations to RF front-end of beam length measurement. We download S-parameter of Mini corporation devices, and use them in simulation. As for frequency mixers, we set their parameters as usual, their conversion loss is 6dB, noise figure 5dB, gain compression power +20dBm. As for amplifier, we set their nose figure as 1.2db, their gain compression power +20dBm. And we design systemic thermal noise as 100 nV. Because the experiment we do is base on real device parameter, this simulation is credible.

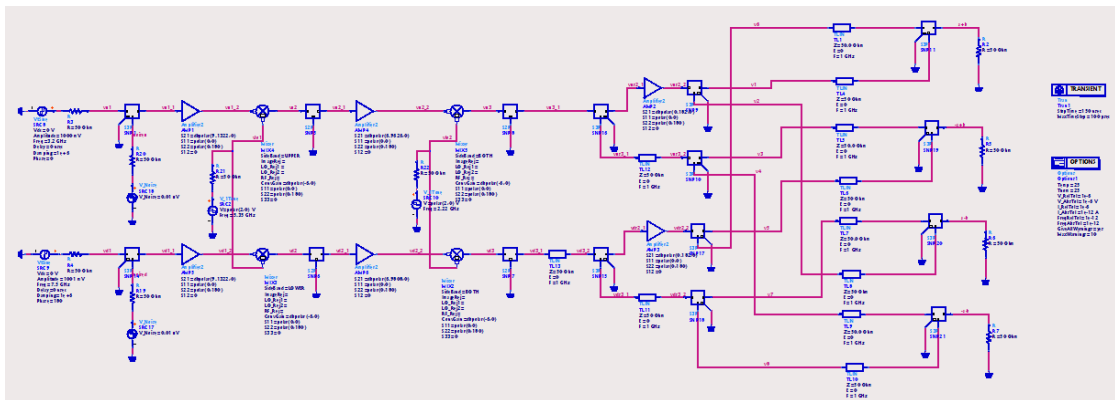

Fig 6, ADS simulation

In this ADS simulation, we use down-conversion twice and adopt the second structure of subtraction and summation. We change the magnitude differential of $V_2$ and $V_1$, and test the linearity between $\Delta V_2$ and their differential.

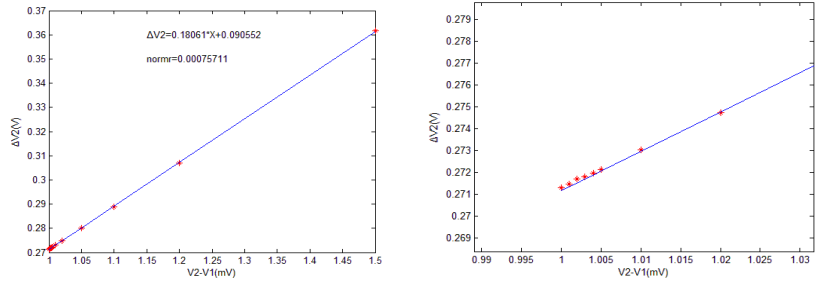

Fig 7.1 and Fig 7.2, the linearity between $\Delta V_2$ and $V_2$-$V_1$

From Fig 7.1 and Fig 7.2 we can say, there is good linearity between $\Delta V_2$ and the differential of $V_2$ and $V_1$. The magnitude differential of $V_2$ and $V_1$ is from 1.001mV to1.5mV, that is to say, $V_2$-$V_1$=1‰$V_1$ is ensured (the magnitude of $V_1$is 1V). Upon this while differential is 0.001mV, there still exist good linearity with $\Delta V_2$. The signal-to-noise ratio of $\Delta V_2$ could reach above 60dB.

CBLM measurement method requires 112dB signal-to-noise ratio, but signal electronic part is hard to make it come true. ADS simulation prove what we design can satisfy this demand, that gives us courage to use electronic method to do this beam length measurement.

**6 Conclusion**

The author has raised a smart method to overcome electronic difficulty. Making subtraction and summation to two channel of signals before inputted to ADC, and this highly lower ADC's demand. We also find a good structure to fulfil this operating at the same time, which can deal with signal coupling effectively.

We adopt ADS software to do simulation to RF front-end part, and in this simulation Mini corporation device are used and other reasonable parameters are set. Simulation has proven the RF front-end we design has signal-to-noise of above 112dB, which can satisfy CBLM systemic require. Above all, electronic method based on CBLM enables to meter sub-picosecond bunch length with good resolution.